\documentclass[journal]{IEEEtran}

\pdfoutput=1 
  \usepackage[pdftex]{graphicx}
   \DeclareGraphicsExtensions{.pdf}
\begin{document}
%
\title{On the correction of anomalous phase oscillation in entanglement witnesses using quantum neural networks}

%
%
%


\author{E.C.~Behrman,
        R.E.F.~Bonde,
        J.E.~Steck,
        and~J.F.~Behrman 
\thanks{E.C. Behrman and R.E.F. Bonde are with the Department
of Mathematics and Physics, Wichita State University, Wichita,
KS, 67260 USA e-mail: elizabeth.behrman@wichita.edu}
\thanks{J.E. Steck is with the Department
of Aerospace Engineering, Wichita State University, Wichita,
KS, 67260 USA e-mail: james.steck@wichita.edu}
\thanks{J.F. Behrman is with the Department
of Physics, Harvard University, Cambridge,
MA, 02138 USA e-mail: jbehrman@college.harvard.edu}
}

%
%

\markboth{Journal of \LaTeX\ Class Files,~Vol.~6, No.~1, January~2007}%
{Shell \MakeLowercase{\textit{et al.}}: Bare Demo of IEEEtran.cls for Journals}
%



\maketitle

\begin{abstract}
Entanglement of a quantum system depends upon relative phase in complicated ways, which no single measurement can reflect. Because of this, entanglement witnesses are necessarily limited in applicability and/or utility. We propose here a solution to the problem using quantum neural networks. A quantum system contains the information of its entanglement; thus, if we are clever, we can extract that information efficiently. As proof of concept, we show how this can be done for the case of pure states of a two-qubit system, using an entanglement indicator corrected for the anomalous phase oscillation. Both the entanglement indicator and the phase correction are calculated by the quantum system itself acting as a neural network. 
\end{abstract}



%
\IEEEpeerreviewmaketitle

\section{Introduction}
%
%
%
%

The scientific process asks questions of the universe. Much of the education of a scientist consists of learning, painstakingly, \underline{how} to ask those questions. If the question is ill-posed, the answer will be ambiguous or even unintelligible; even if it is well-posed, the question needs to be the one whose answer is pertinent to whatever problem it is we wish to solve. Simulations are one good route to finding out about the clarity and worth of our questions, since variables can be easily controlled and extraneousness determined.

Quantum imformation is the systematic use of the fundamental quantum mechanical nature of the universe to do calculations that are either very difficult or even impossible to do with a classical computer. Exactly what this allows us to do, and how we can make use of that nature, turns out to be a very difficult problem. The quantum system itself contains that information. But how do we exploit those abilities?  How do we even find out what they are? 

In this paper we use a Quantum Neural Network (QNN), which necessarily complex-valued, to calculate the essential attributes of the quantum information. While this is done in simulation, the method, as outlined, could be implemented experimentally to create a quantum computer, programmed through neural network training, to calculate its own entanglement and phase.

\section{Entanglement}

Determination of entanglement is one very good example. This is an important question because entanglement is what allows us to \underline{do} quantum computations \cite{genentref}; it is an essentially \underline{quantum} question since no classical system can have quantum correlations; thus, it is an excellent testbed for this general problem.  The quantum system which is our quantum computer ``knows'' what its entanglement is, but how do we ask the question? And how do we know if we have the right answer?

Given the density matrix, entanglement can actually be calculated for the two-qubit system. For definiteness, we choose the ``entanglement of formation'' $E_{F}$ \cite{wootters}, which is the number of pure singlets necessary to create a given entangled state.  Of course, we may not even know the density matrix. We could determine the density matrix  through quantum tomography \cite{tomography}, but the number of measurements necessary goes like $2^{2N}$, where $N$ is the number of qubits. Thus, as the system grows this method becomes prohibitive. Is there instead a relatively cheap way to determine the entanglement? 

A number of researchers  (see, {\it e.g.}, \cite{mintert, toth}) have devised single measurement  ``entanglement witnesses'', to indicate the presence or absence of entanglement. An entanglement witness, $W$, is a Hermitian operator such that $tr(W \rho)\ge 0$ for a density matrix $\rho$ representing a fully separable (i.e., unentangled) state. Most of them are measures of ``closeness'' to a single, entangled state, and fail for other kinds of entangled states. This is easiest to understand by considering the so-called ``Bell'' states: 
\begin{eqnarray}
|\Phi_{+}(\theta_{1})\rangle = \frac{1}{\sqrt{2}}[|00\rangle+e^{i \theta_{1}}|11\rangle]\\ \nonumber 
|\Phi_{-}(\theta_{2})\rangle = \frac{1}{\sqrt{2}}[|00\rangle-e^{i \theta_{2}}|11\rangle]\\ \nonumber 
|\Psi_{+}(\theta_{3})\rangle = \frac{1}{\sqrt{2}}[|01\rangle+e^{i \theta_{3}}|10\rangle]\\ \nonumber 
|\Psi_{-}(\theta_{4})\rangle = \frac{1}{\sqrt{2}}[|01\rangle-e^{i \theta_{4}}|10\rangle
\label{Bellbasis}
\end{eqnarray}
For $\{\theta_{i}\} = 0$, this is a commonly used basis for the two-qubit system, the ``Bell basis'' (it is an orthonormal set). Each of these states is fully entangled, $E_{F} = 1$  (i.e., each could be created using exactly one singlet), independent of the angle $\theta_{i}$.

\begin{figure}   
\centering
\includegraphics[width=2.5in]{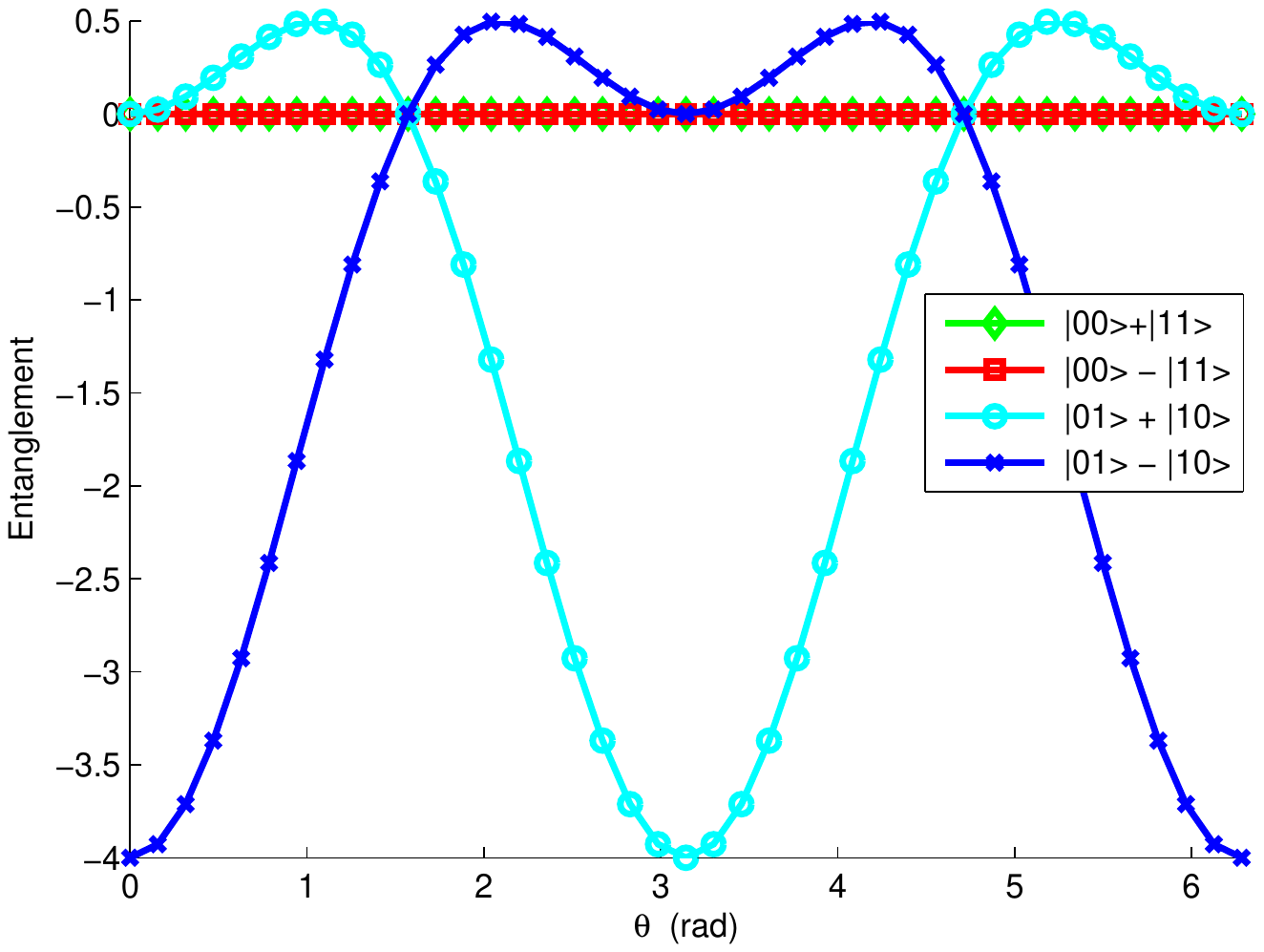}
\caption{The Mintert Witness, $W_{M}$, set for closeness to $|\Psi_{-}(\theta_{4}=0)\rangle$, tested on each of the Bell basis states as functions of phase offset $\theta$.  Both $|\Phi_{+}\rangle$ and $|\Phi_{-}\rangle$ are calculated to have zero entanglement, while $|\Psi_{+}\rangle$ and $|\Psi_{-}\rangle$ oscillate, showing entanglement for half the range of $\theta$. Note that the \underline{correct} answer is that each state is fully entangled for all $\theta$ values (i.e., the witness should be negative everywhere.) This oscillation is what we are calling ``anomalous oscillation'' in the entanglement witness.   \label{Mintert1fig}}
\end{figure}

A witness can be devised that measures closeness to any one of these states. For example, the Mintert witness \cite{mintert} is given by $W_{M} =-4 tr(\rho \otimes \rho V)$, where $V = P_{-}^{(1)} \otimes (P_{-}^{(2)}  - P_{+}^{(2)})$, $P_{-}$ is the projector onto the antisymmetric states, and $P_{+}$ is the projector onto the symmetric states.  As proposed this witness is set up to work well for states close to the singlet (antisymmetric) state, $|\Psi_{-}(\theta_{4}=0)\rangle$, so it does not detect entanglement in either $|\Phi_{+}\rangle $ or $|\Phi_{-}\rangle $. Worse, the witness increasingly fails to detect entanglement in  $|\Psi_{-}(\theta_{4})\rangle$  as $\theta_{4}$ increases. See Figure \ref{Mintert1fig}, which shows $W_{M}$ (set for closeness to the EPR singlet, $|\Psi_{-}(\theta_{4}=0)\rangle$) , calculated for each of the Bell states as functions of their respective $\theta$. Entanglement is indicated by negativity of the result. The straight line showing the constant result of zero shows that this witness predicts no entanglement for either $ |\Phi_{+}\rangle$ or $|\Phi_{-}\rangle$. The two oscillatory curves correspond to the $|\Psi_{+}\rangle$ and $|\Psi_{-}\rangle$ states. At $\theta = \pi/2$  each becomes the other; thus, at that point, $|\Psi_{-}\rangle$ is predicted to lose entanglement and $|\Psi_{+}\rangle$  to become entangled. 

\begin{figure}   
\centering
\includegraphics[width=2.5in]{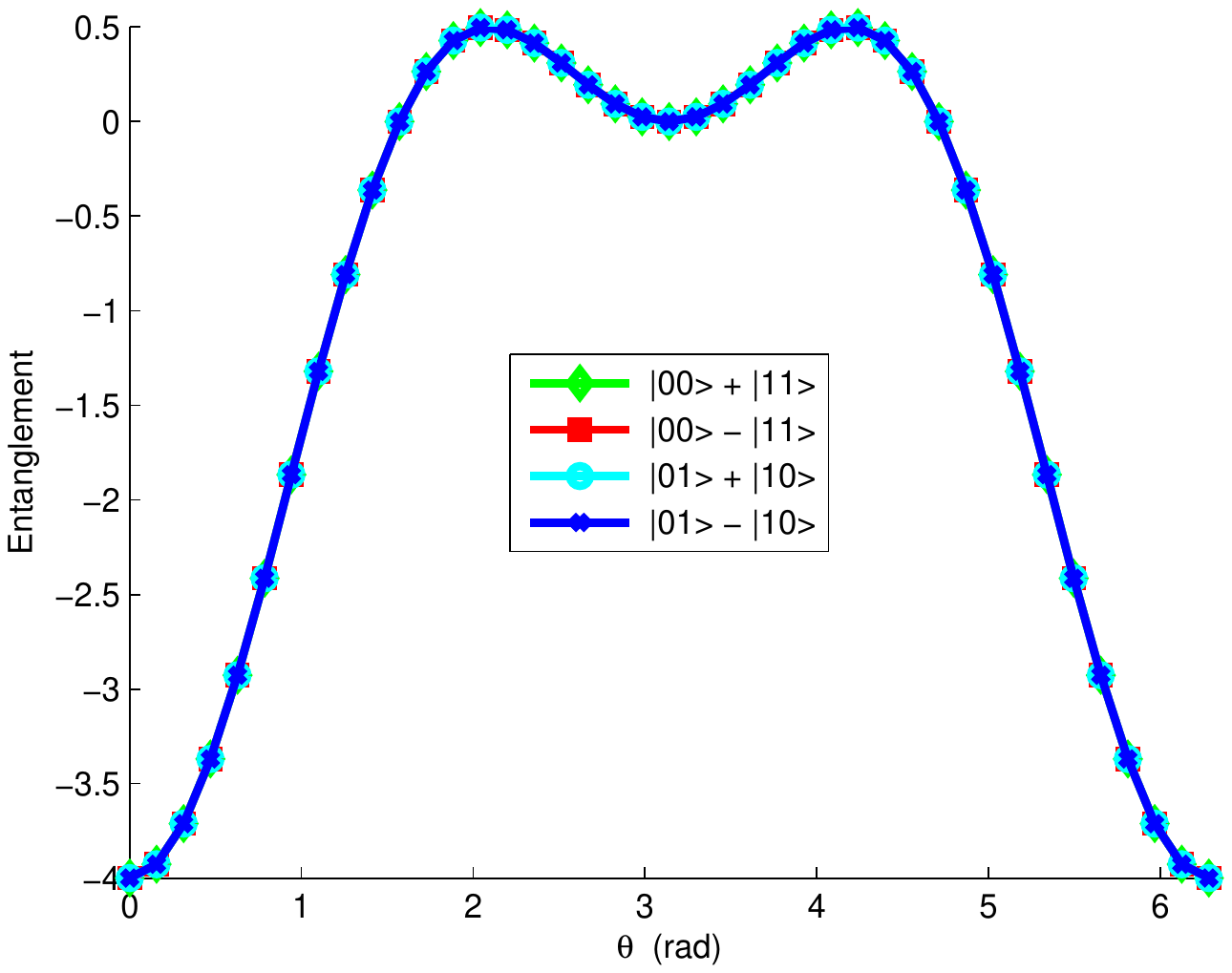}
\caption{The Mintert Witness, $W_{M}$, set for closeness to each of the Bell basis states (at $\theta = 0$) in turn, and tested as functions of phase offset $\theta$.  All show exactly the same oscillatory behavior. Again, the correct answer is that each state is fully entangled for all $\theta$ values (the witness should be negative everywhere.) \label{Mintert2fig}}
\end{figure}

Witnesses can be ``reset'' so that they will work for other members of the Bell basis; for this witness, all that is necessary is to change the projection operators. See Figure \ref{Mintert2fig}, which shows results of this procedure: all the predictions now lie atop each other (as they must do by symmetry.) So, if you know which entangled state your unknown state is close to, you can use a witness which will correctly predict its entanglement. However this creates an obvious dilemma: {\it you must have knowledge of the state before you do the measurement to gain knowledge of the state.} 

All single measurement witnesses must show this anomalous oscillation \cite{eb08}, shown in Figures \ref{Mintert1fig} and \ref{Mintert2fig}. Clearly, it would be of value to devise a corrector to this oscillation, so that the witness would then correctly predict that all the states of Equation 1 are fully entangled. With such a method, we could thereby determine whether a general, unknown, input state was entangled, and perhaps even the degree of entanglement. We propose here such a method, using a quantum neural network. 

\section{Quantum neural network \label{qnn}}

In previous work we have shown that a quantum system can be trained to deal with both parts of the problem: that it can be trained to output an entanglement indicator\cite{eb08, nabic, eb12} or a phase indicator \cite{singapore}. Briefly, the details are as follows.  

For the 2-qubit system, we write for the Hamiltonian:
\begin{equation}
H = K_{A} \sigma_{xA} + K_{B} \sigma_{xB} + \varepsilon_{A} \sigma_{zA} + \varepsilon_{B} \sigma_{zB} + \zeta \sigma_{zA} \sigma_{zB} 
\label{hamiltonian}
\end{equation}
where $\{ \sigma \}$ are the Pauli operators corresponding to each of the two qubits, A and B, $K_{A}$ and $K_{B}$ are the tunneling amplitudes, $\varepsilon_{A}$ and $\varepsilon_{B}$  are the biases, and $\zeta$ the qubit-qubit coupling. (Generalization to an N-qubit system is possible \cite{nabic, eb12} but will not be considered here.) The time evolution of the system is given by the  Schr\"{o}dinger equation:
\begin{equation}
\frac{d \rho}{dt} = \frac{1}{i \hbar}[H, \rho] 
\label{schr}
\end{equation}
where $\rho$ is the density matrix and $H$ is the Hamiltonian. The parameters $\{ K,\varepsilon,\zeta \}$ determine the time evolution of the system in the sense that, if one or more of them is changed, the way a given state will evolve in time will also change. This is the basis for using our quantum system as a neural network. The role of the ``weights'' of the network is played by the parameters of the Hamiltonian, $\{ K,\varepsilon,\zeta  \}$, all of which we take to be  experimentally adjustable as functions of time (see, {\it e.g.}, \cite{yamamoto}, for the case of SQuID charge qubits.) By adjusting the parameters using a neural  learning algorithm we can train the system to evolve in time to a set of chosen target outputs at the final time $t_{f}$, in response to a corresponding (one-to-one) set of given inputs. Because the time evolution is quantum mechanical (and, we assume, coherent), a quantum mechanical function can be mapped to an observable of the system's final state, a measurement made at the final time $t_{f}$. The time evolution of the quantum system is calculated by integrating the Schr\"{o}dinger equation numerically in MATLAB Simulink, using ODE4 (Runge-Kutta), with a fixed integration step size of 0.05 ns \cite{matlab}. The system was initialized in each input state in the training set, in turn, then allowed to evolve for 190 ns. A measurement is then made at the final time; this is the ``output'' of the network. An error, $target-output$, is calculated, and the parameters are adjusted slightly to reduce the error. This is repeated for each $(input,target)$ pair multiple times until the calculation converges on parameters that work well for the entire training set.  Complete details, including a derivation of the quantum dynamic learning paradigm using backpropagation \cite{lecun} in time \cite{werbos}, are given in \cite{eb08} and in \cite{singapore}.

\subsection{QNN entanglement indicator}

For the entanglement indicator, we chose as our output the qubit-qubit correlation function \cite{recent} evaluated at the final time, $\langle \sigma_{zA}(t_{f}) \sigma_{zB}(t_{f}) \rangle^{2}$, and trained the indicator using a training set of four (one fully entangled state, two product (unentangled) states, and one partially entangled state.) That is, the input,output) pairs are: 

\begin{eqnarray}
input =  |\Psi(0)\rangle \\ \nonumber
output = \langle \sigma_{zA}(t_{f}) \sigma_{zB}(t_{f}) \rangle ^{2} \rightarrow target 
\end{eqnarray} 

\noindent with prepared input states at zero time, and corresponding targets, given in Table \ref{enttraining}, which also shows the trained values and the entanglement of formation, calculated using the analytic formula \cite{wootters} for comparison. Note that the QNN indicator systematically underestimates $E_{F}$ for partially entangled states; this is because we found through simulation that the net naturally trained to the target value of 0.44. See \cite{eb08} for details. That is, we seek here not exact \underline{agreement} with $E_{F}$ (in which case we would train the state $\frac{1}{\sqrt{3}}(|00\rangle + |01\rangle + |10\rangle)$ to a target value of 0.55), but a robust and internally  self consistent measure, which we would hope would track well with an analytic measure like $E_{F}$.

\begin{table}[!t]
\caption{Training data for QNN entanglement witness.}
\label{enttraining}
\begin{tabular}{l l l l}\\
\hline
Input state $|\Psi(0)\rangle$  & Target  & Trained & $E_{F}$ \\
\hline
$\frac{1}{\sqrt{2}}(|00\rangle + |11\rangle)$ & 1.0 & 0.998 & 1.0 \\
$\frac{1}{2}(|00\rangle + |01\rangle + |10\rangle + |11\rangle)$    & 0.0  &  $1.2 \times 10^{-5}$ & 0.0 \\
$\frac{1}{\sqrt{1.25}}(0.5|10\rangle + |11\rangle) $ & 0.0  &  $1.8 \times 10^{-4}$  & 0.0 \\
$\frac{1}{\sqrt{3}}(|00\rangle + |01\rangle + |10\rangle)$   & 0.44  & 0.44 & 0.55 \\
\hline
RMS    &   { }  & $4.4 \times 10^{-6}$ & { }  \\
Epochs  &  { }   &  100  &  {} \\
\hline \\
\end{tabular}
\end{table}

\begin{figure}   
\centering
\includegraphics[width=2.5in]{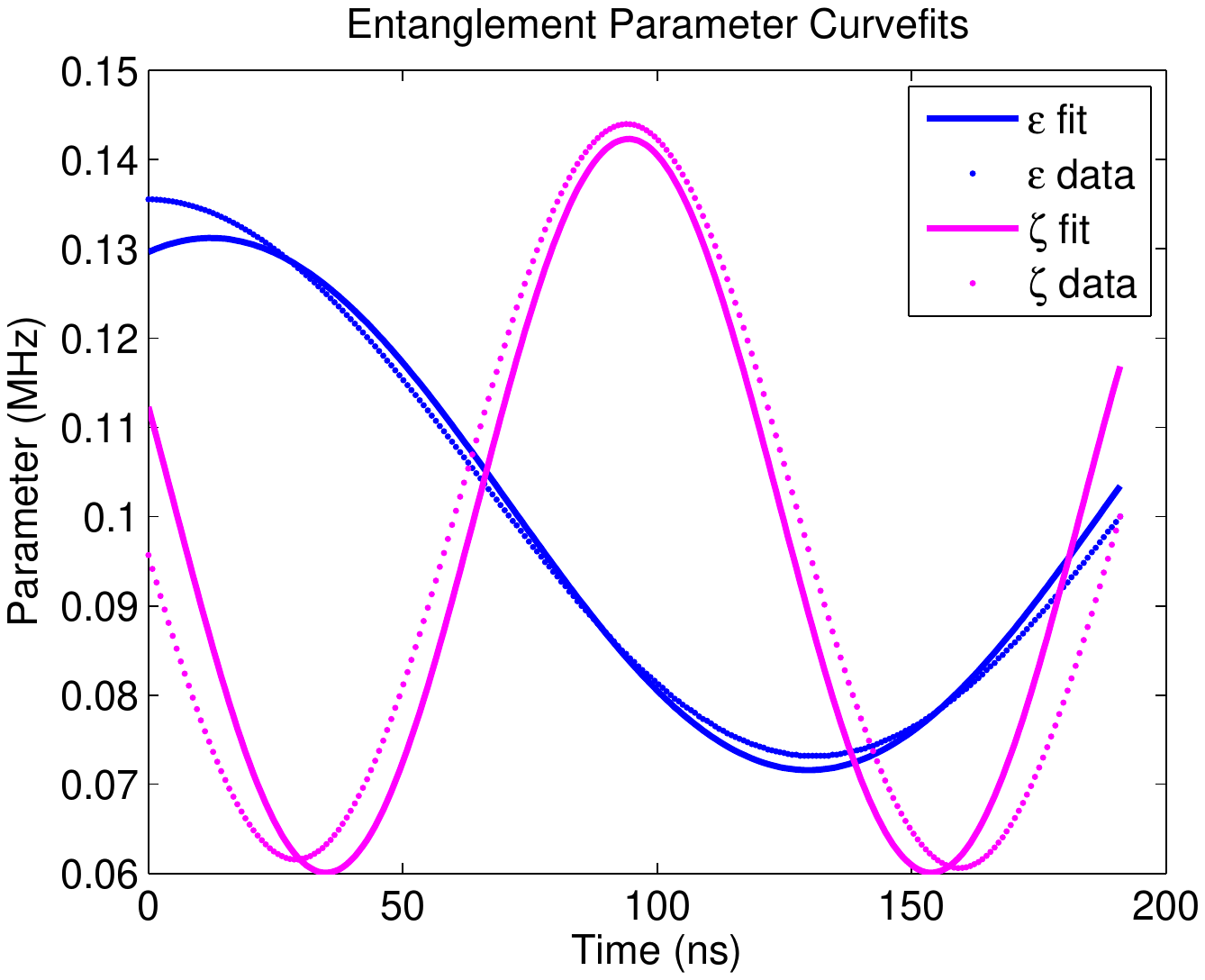}
\caption{The functions  $\epsilon_{A}=\epsilon_{B}$ and $\zeta$, as functions of time, as trained for the entanglement indicator, and plotted with a single frequency Fourier fit.  Each was started out (pre-training values) as a constant function:   $\epsilon_{A}=\epsilon_{B}=\zeta=10^{-4} GHz$.\label{paramEfig} }
\end{figure}

\begin{figure}   
\centering
\includegraphics[width=2.5in]{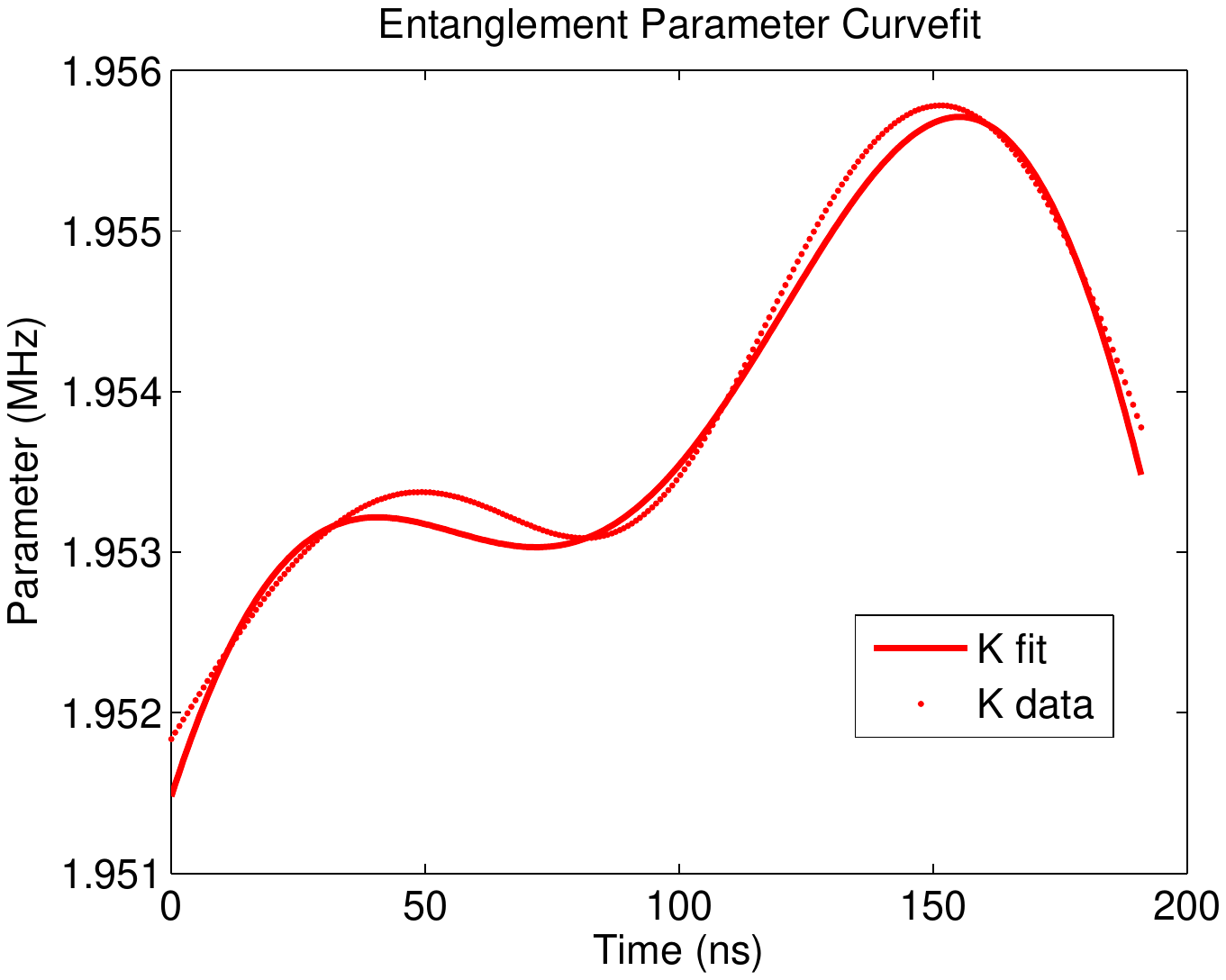}
\caption{The function $K_{A}=K_{B}$ as a function of time, as trained for the entanglement indicator, and plotted with a two-frequency Fourier fit. $K$ was started out (pre-training values) as a constant function:  $K_{A} = K_{B}= 1.875 \times 10^{-3} GHz$. \label{paramKEfig} }
\end{figure}

Note that the results in Table \ref{enttraining} are retrained from our previous work \cite{eb08}, this time using continuously varying functions rather than functions piecewise constant in time; training was, as might be expected, much more rapid, and the trained parameter functions more symmetric: $K_{A}$ and $K_{B}$ lie right on top of each other, as do $\epsilon_{A}$ and $\epsilon_{B}$. Both $\epsilon$ and $\zeta$ were simple oscillatory functions, of only a single frequency; however, $K$ seems to exhibit two frequencies. We curvefit each function to a Fourier series. The agreement was very good. Coefficients for the fits are shown in Table \ref{entparamfit}. The trained functions, with their curvefits, are shown in Figures \ref{paramEfig} and \ref{paramKEfig}.

\begin{table}[!t]
\caption{Curvefit coefficients for parameter functions $K$, $\epsilon$, $\zeta$, for QNN entanglement witness.}
\label{entparamfit}
$f(t) =  a_{0} + a_{1}cos(\omega t) + b_{1}sin(\omega t) + a_{2}cos(2\omega t) +  b_{2}sin(2\omega t) $ \\
\begin{tabular}{l l l l }\\
coefficient  & K & epsilon & zeta \\
\hline
 $a_{0}$  &   $  0.0019495 $ & $1.014\times 10^{-4}$ & $1.012 \times 10^{-4}$ \\
 $a_{1}$  &   $-1.002\times 10^{-6}$ &$ 2.824\times 10^{-5}$ & $ 1.109\times 10^{-5}$ \\
 $b_{1}$  &   $ 6.868\times 10^{-6}$ & $9.577\times 10^{-6}$  & $ -3.96\times 10^{-5} $ \\
 $a_{2}$  &  $  2.981\times 10^{-6}$ &  ---  &  ---  \\
 $b_{2}$  &  $-4.562\times 10^{-7}$ &   ---  &  ---  \\
 $\omega$   &   $  0.01645  $   &$ 0.02674 $ &$ 0.05282 $ \\
\hline
RMS    &  $1.069\times 10^{-7} $ & $1.88\times 10^{-6}$ &  $7.982\times 10^{-6}$  \\   
\hline \\
\end{tabular}
\end{table}

\begin{figure}[!t]   
\centering
\includegraphics[width=2.5in]{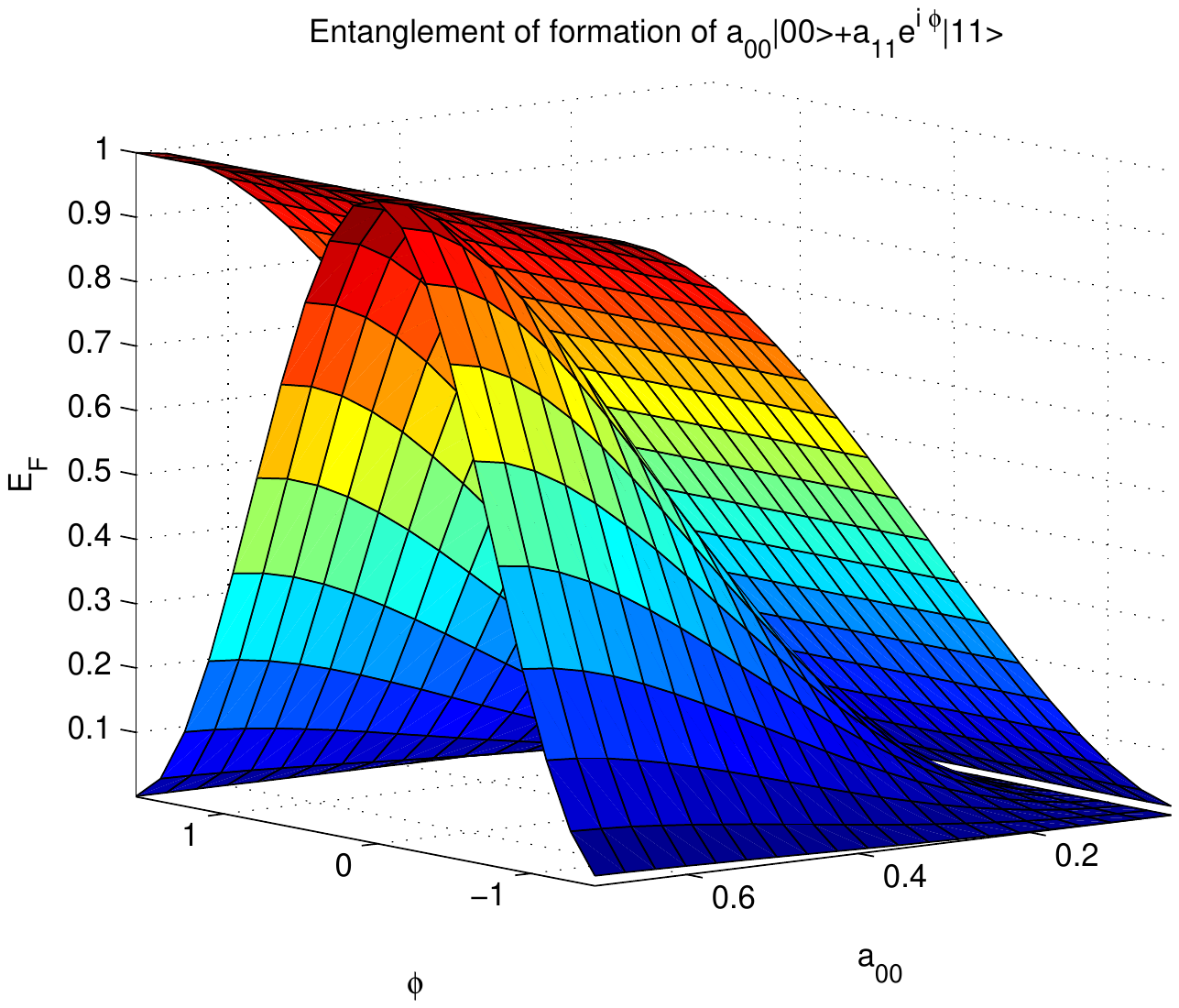}
\caption{QNN entanglement indicator for states of the form $a_{00}|00\rangle + a_{11}e^{i \phi}|11\rangle$, where $a_{00}$ and $a_{11}$ are both positive real, as a function of both relative magnitude and $\phi$, and compared with the entanglement of formation. }\label{QNNBellfig}
\end{figure}

\begin{figure}[!t]   
\centering
\includegraphics[width=2.5in]{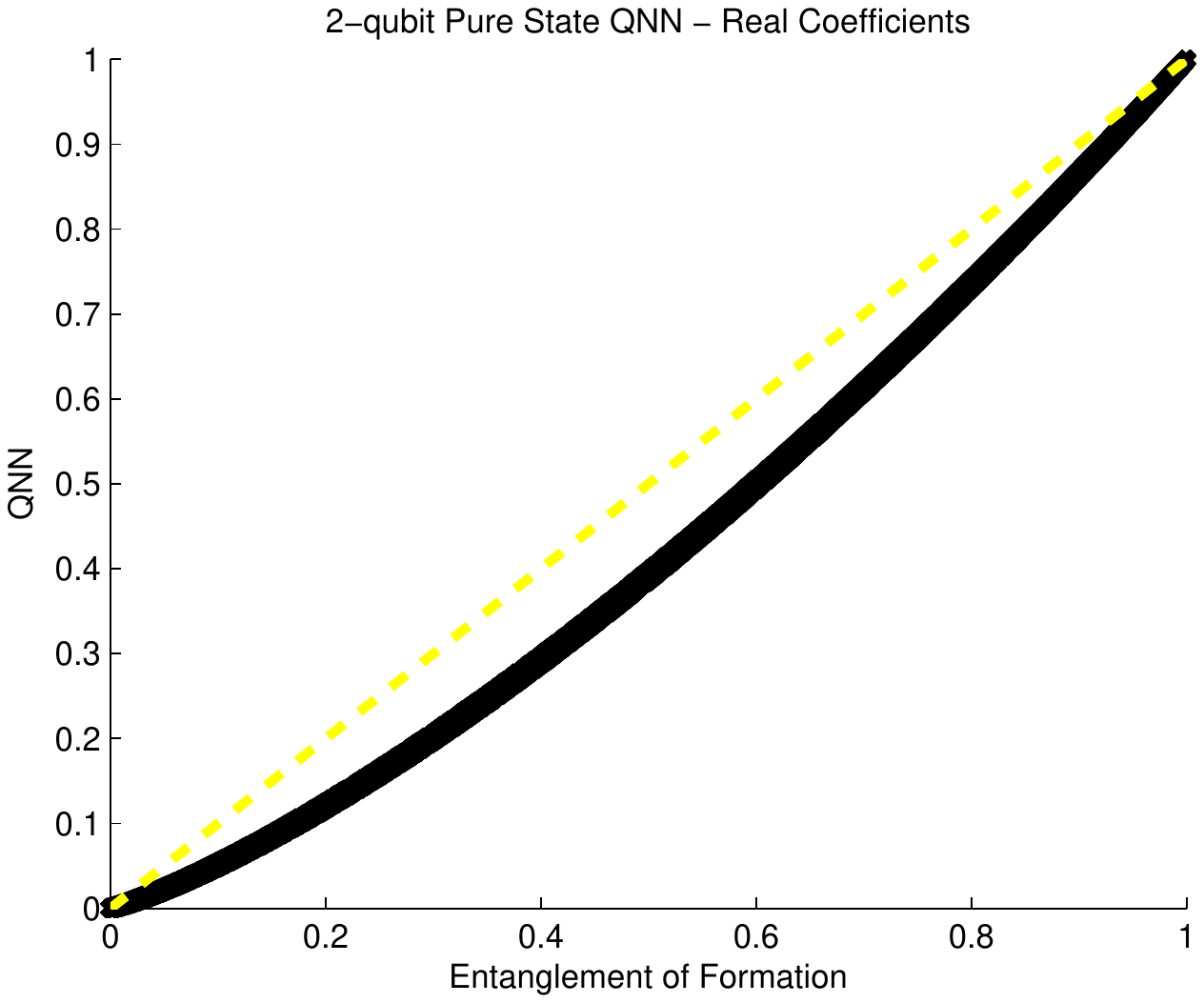}
\caption{QNN entanglement for 50,000 randomly generated pure states of the form $a_{00}|00\rangle + a_{01}|01\rangle + a_{10}|10\rangle + a_{11}|11\rangle$, where $a_{00}$, $a_{01}$, $a_{10}$,and $a_{11}$ are all real, as a function of the entanglement of formation. Points lying along the dashed yellow line are states for which the entanglement predicted by the QNN witness exactly matches the entanglement of formation. }\label{realfig}
\end{figure}

\begin{figure}   
\centering
\includegraphics[width=2.5in]{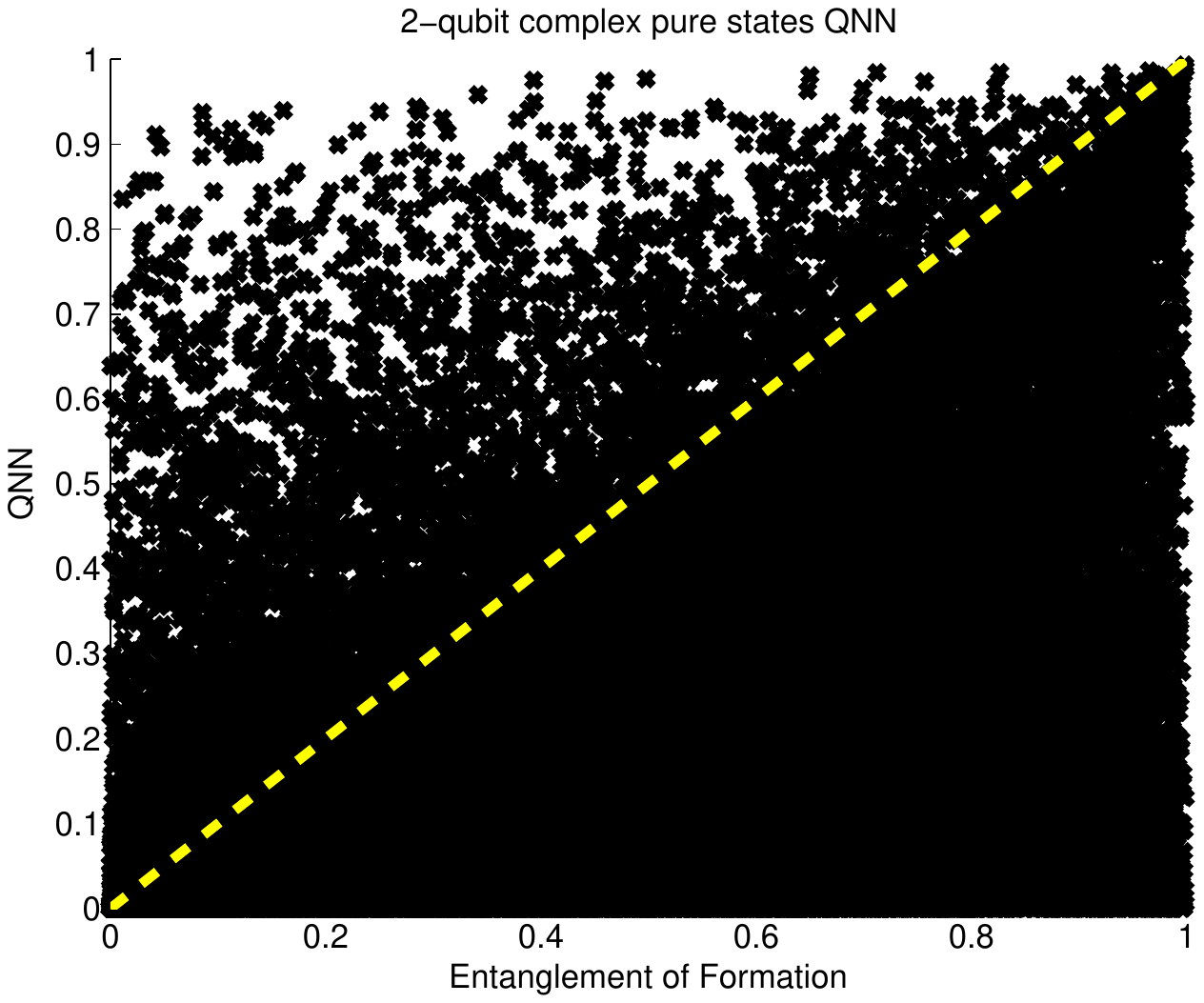}
\caption{As in Figure \ref{realfig}, but with complex coefficients. \label{cxcoeffig}}
\end{figure}

This indicator gives good results for large classes of input states, including both pure and mixed states, and is not restricted to states  ``close'' to any particular state. We have also \cite{nabic,eb12} extended our work to the 3-, 4-, and 5-qubit cases, and found that as the size of the system grows, the amount of additional training necessary diminishes; thus, our method may be very practical for use on large computational systems. But as it is a single measurement, we too get anomalous oscillation. See Figure \ref{QNNBellfig}, which shows that the QNN indicator tracks very well with $E_{F}$ for states of the form $a_{00}|00\rangle + a_{11}e^{i \phi}|11\rangle$ as a function of $a_{00}$, but predicts, incorrectly, that the entanglement is a function of $\phi$. Another way of looking at this is shown in Figure \ref{realfig} and Figure \ref{cxcoeffig}. If we confine our testing to states with only real coefficients, the QNN gives an excellent approximation to the entanglement of formation. Figure \ref{realfig} shows a comparison of the QNN entanglement indicator for 50,000 randomly generated states with real coefficients for the 2-qubit system. Agreement is excellent ($45^{\circ}$ yellow line is ideal); however, if we relax the restriction, and include complex coefficients (or, equivalently, nonzero phase offsets), our agreement becomes quite bad (see Figure \ref{cxcoeffig}.) So, we need to know the phase offset(s) of the system before we can calculate the entanglement.

\subsection{Phase indicator}

\begin{figure}   
\centering
\includegraphics[width=2.5in]{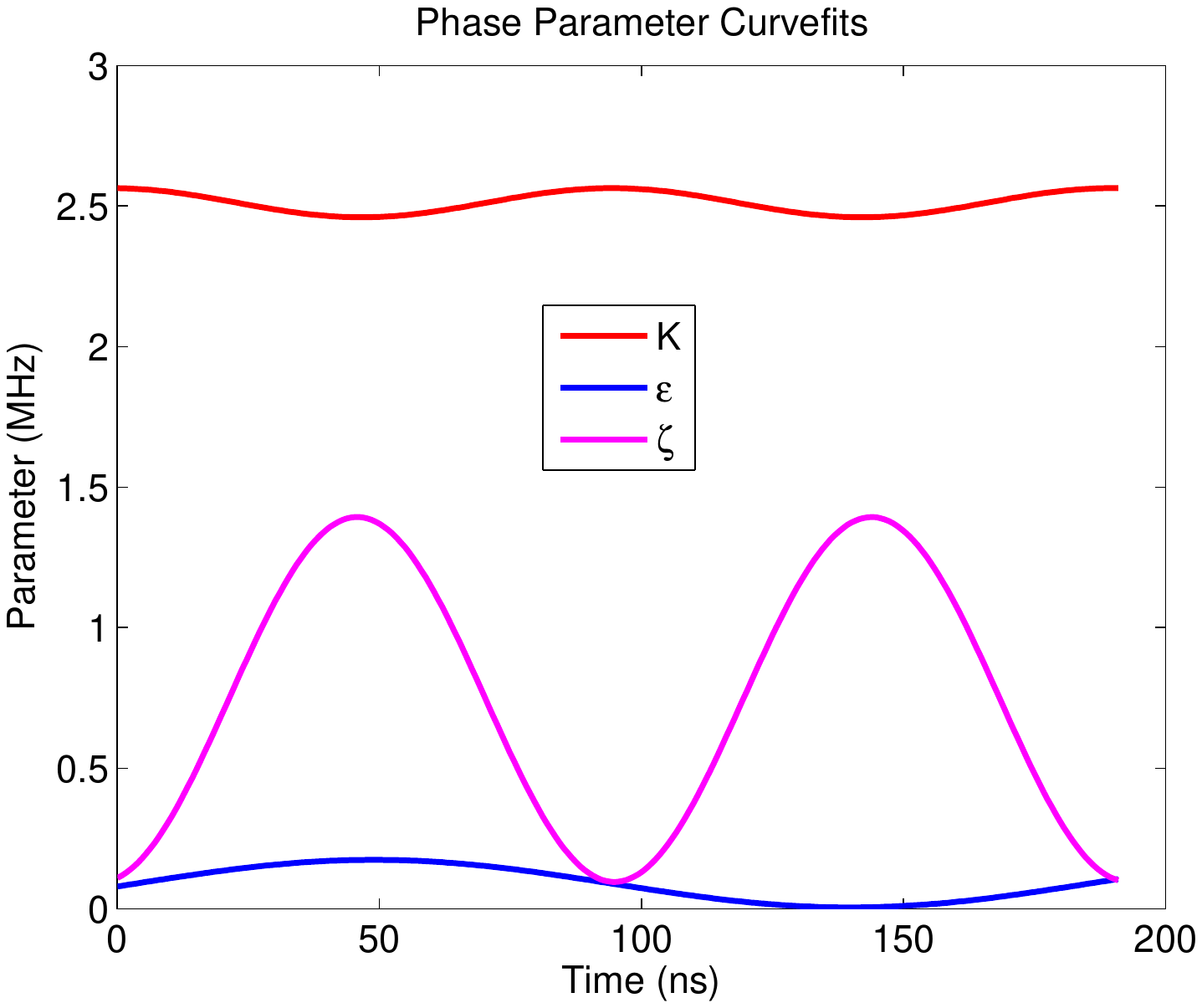}
\caption{The functions $K_{A}=K_{B}$, $\epsilon_{A}=\epsilon_{B}$, and $\zeta$, as functions of time, as trained for the phase offsets, and plotted with curvefits to single oscillatory functions, as explained in the text. On the scale of the graph the curvefits are indistinguishable from the plotted points of the trained values. \label{paramphasefitfig}}
\end{figure}

Since the phase offset is yet another quality of the system, the system should, itself, be able to supply us with information about its own phase. And so it can. We choose the ``charge'' basis, in which a general pure state can be written

\begin{eqnarray}
|\Psi(0)\rangle=a_{00}|00\rangle+a_{01}e^{i\xi}|01\rangle+a_{10}e^{i \theta}|10\rangle \\ \nonumber +a_{11}e^{i \phi}|11\rangle
\end{eqnarray}
where normalization requires that 
\begin{equation}
\sqrt{a_{00}^{2}+a_{01}^{2}+a_{10}^{2}+a_{11}^{2}}=1
\end{equation}
Since an \underline{overall} phase is physically meaningless we may take out any overall phase factor; that is, without loss of generality we may take the coefficient of the $|00\rangle$ basis state to be real. We then write each of the other coefficients as its magnitude times a phase factor; thus, each $a_{nm}$ will be a real number, and the phase factor, if any, will be written in explicitly.  There are, thus, three independent phases for the two-qubit system. Our training set consisted of (input,output) pairs of the form

\begin{eqnarray}
input =  |\Psi(0)\rangle= \frac{1}{\sqrt{2}}(|00\rangle + e^{i \phi}|11\rangle) \\ \nonumber
output = |\langle11|\Psi(t_{f})\rangle|^{2} \rightarrow target = \cos^{2}(\phi/2)
\label{phIO}
\end{eqnarray} 

\noindent with 11 different values of $\phi$ from $-\pi$ to $\pi$. Training was rapid and good. Figure \ref{paramphasefitfig} shows the trained parameter functions for the phase indicator. $K_{A}$ and $K_{B}$ lie right on top of each other, as do $\epsilon_{A}$ and $\epsilon_{B}$.  Each was started out (pre-training values) as a constant function:  $K_{A} = K_{B}= 2.5 \times 10^{-3}$GHz, and $\epsilon_{A}=\epsilon_{B}=\zeta=10^{-4}$GHz. Each trained rapidly to what seemed obviously to be a single oscillatory function. We therefore curvefit each function to a single term in the Fourier series.  Coefficients for the fits are shown in Table \ref{phaseparamfit}. 

\begin{table}[!t]
\caption{Curvefit coefficients for parameter functions $K$, $\epsilon$, $\zeta$, for QNN phase indicator.}
\label{phaseparamfit}
$f(t) =  a_{0} + a_{1}cos(\omega t) + b_{1}sin(\omega t) $ \\
\begin{tabular}{l l l l }\\
coefficient  & K & epsilon & zeta \\
\hline
 $a_{0}$  &    $0.002512$ & $8.945\times 10^{-5}$ &$ 7.445\times 10^{-4} $ \\
 $a_{1}$  &   $5.156\times 10^{-5}$ & $-1.005\times 10^{-5}$ & $ -6.346\times 10^{-4} $ \\
 $b_{1}$  &   $ -3.781\times 10^{-6}$ & $ 8.4e-005$  & $ 1.359 \times 10^{-4}$ \\
 $\omega$   &    $   0.0658  $  &$ 0.03454$ & $0.06402$  \\
\hline
RMS    & $ 7.556\times 10^{-6}$  &$ 3.869\times 10^{-6}$ &   $4.468\times 10^{-6}$  \\   
\hline \\
\end{tabular}
\end{table}
 
    

    

    

Agreement is good even with the fitted functions in place of the trained ones; and it is certainly easier to use a fixed analytic function than a large collection of discrete data samples. For testing on more general states than those of Equation 7, though, we needed to expand the target function. 

Because quantum computational power comes from entanglement, it is not surprising that the ability of the net to determine a relative phase between two basis states depends on there being entanglement between those states, and, indeed, the size of the signal diminishes with decreasing entanglement. For the Bell states with unequal magnitude, $ a_{00}|00\rangle+a_{11}e^{i \phi}|11\rangle$ , we found that if we used the target function $2(\frac{1}{2} - a_{00}^{2})^{2} a_{11}^{2} +2 a_{00}a_{11}\cos^{2}(\phi/2)$ we got excellent results. Note that $2 a_{00}a_{11}$, the coefficient in front of the $\cos^{2}(\phi/2)$ term, is the concurrence for this state - a monotonic measure, like $E_{F}$, of the entanglement \cite{wootters}.  Testing on states of the form $a_{00}|00\rangle+a_{01}|01\rangle+a_{11}e^{i \phi}|11\rangle$ was equally good when we again adjusted for the diminished amount of entanglement in the target function. 
 
In fact, the parameter functions $\{K_{A}=K_{B},\epsilon_{A}=\epsilon_{B},\zeta\}$ we found through training only on the eleven states of Equation 7 gave us all three of the phases $\{ \phi, \theta, \xi \}$, with symmetrically adjusted outputs and target functions (e.g., to find $\theta$ we used the output $|\langle10|\Psi(t_{f})\rangle|^{2}$.)
 
\section{Correction of anomalous phase oscillation}

\begin{figure}   
\centering
\includegraphics[width=2.5in]{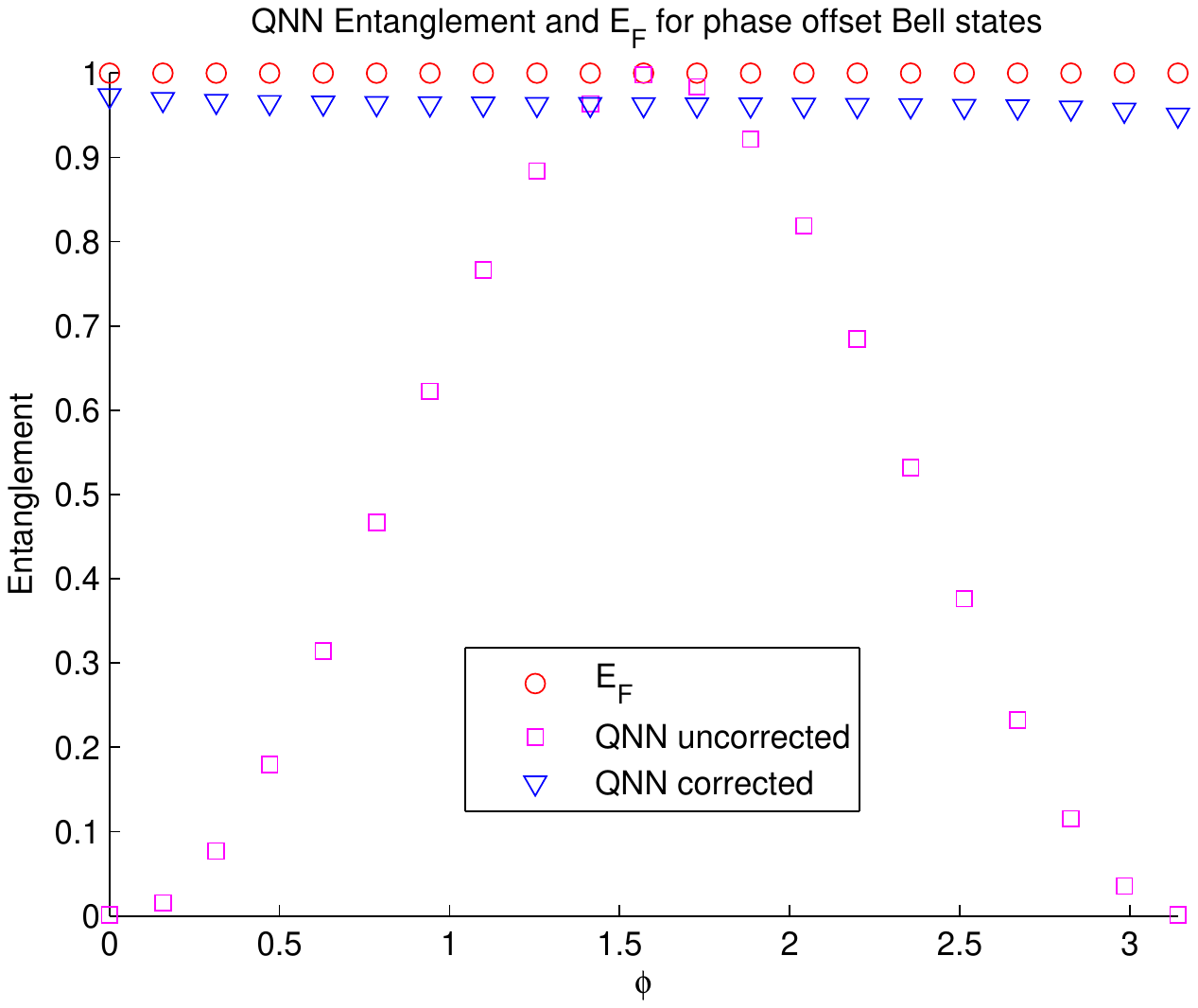}
\caption{The entanglement of $\frac{1}{\sqrt{2}}[|00\rangle + e^{i \phi}|11\rangle]$, as a function of $\phi$, as calculated by the QNN, both with and without phase correction as described in the text, and compared with $E_{F}$ (equal to 1 for all $\phi$.) Results are exactly similar for the $\frac{1}{\sqrt{2}}[|01\rangle + e^{i \theta}|10\rangle]$ states. \label{fixbellfig}}
\end{figure}

\begin{figure}   
\centering
\includegraphics[width=2.5in]{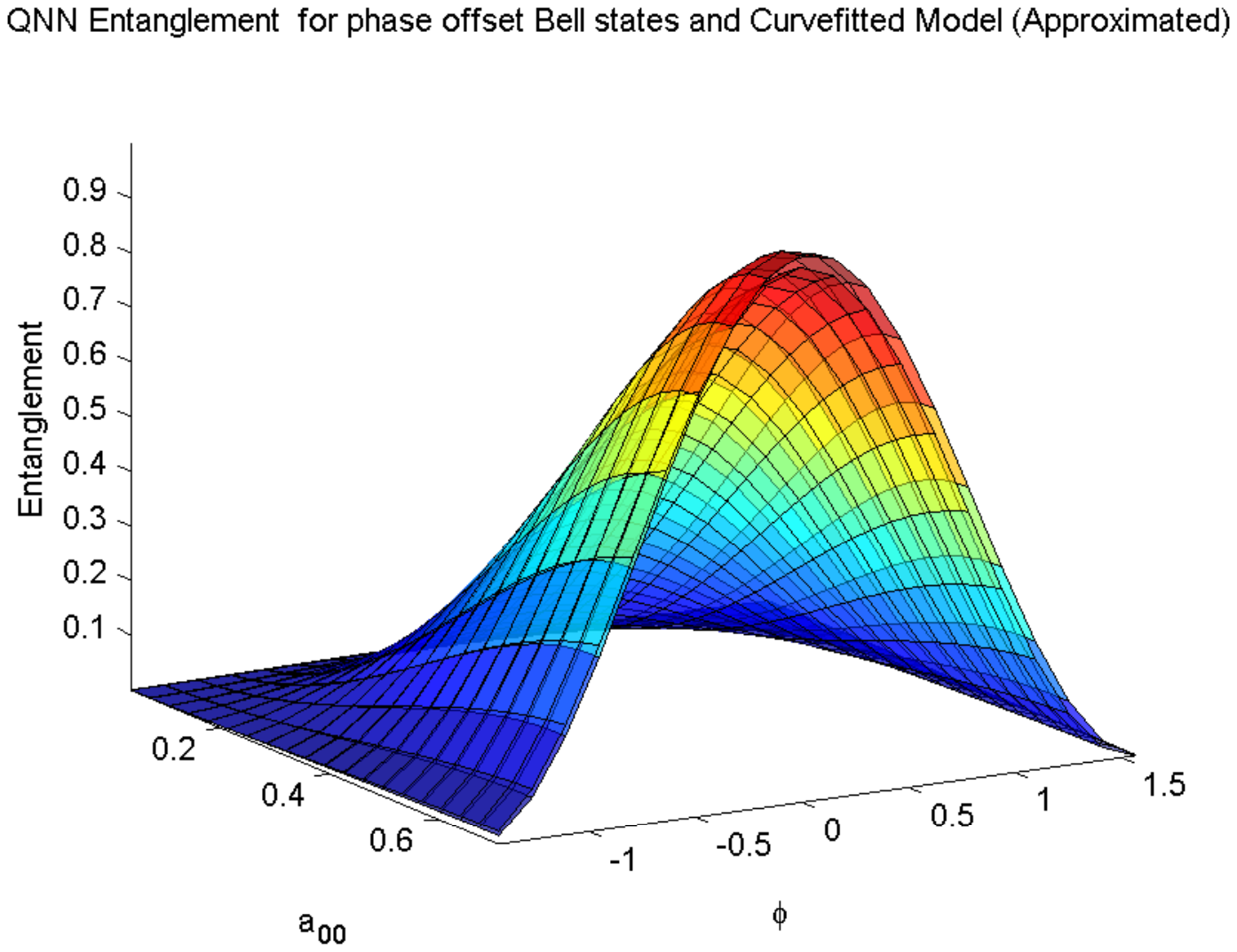}
\caption{The entanglement of $[a_{00}|00\rangle + e^{i \phi}a_{11}|11\rangle]$, as a function of $\phi$ and $a_{00}$, as calculated by the QNN,  and compared with a curvefit to $\sin^{2}(2a_{00})\cos^{2}(\phi)$ (the partially transparent grid, superimposed.) \label{qnn_model_approximation_jan12fig}}
\end{figure}

\begin{figure}   
\centering
\includegraphics[width=2.5in]{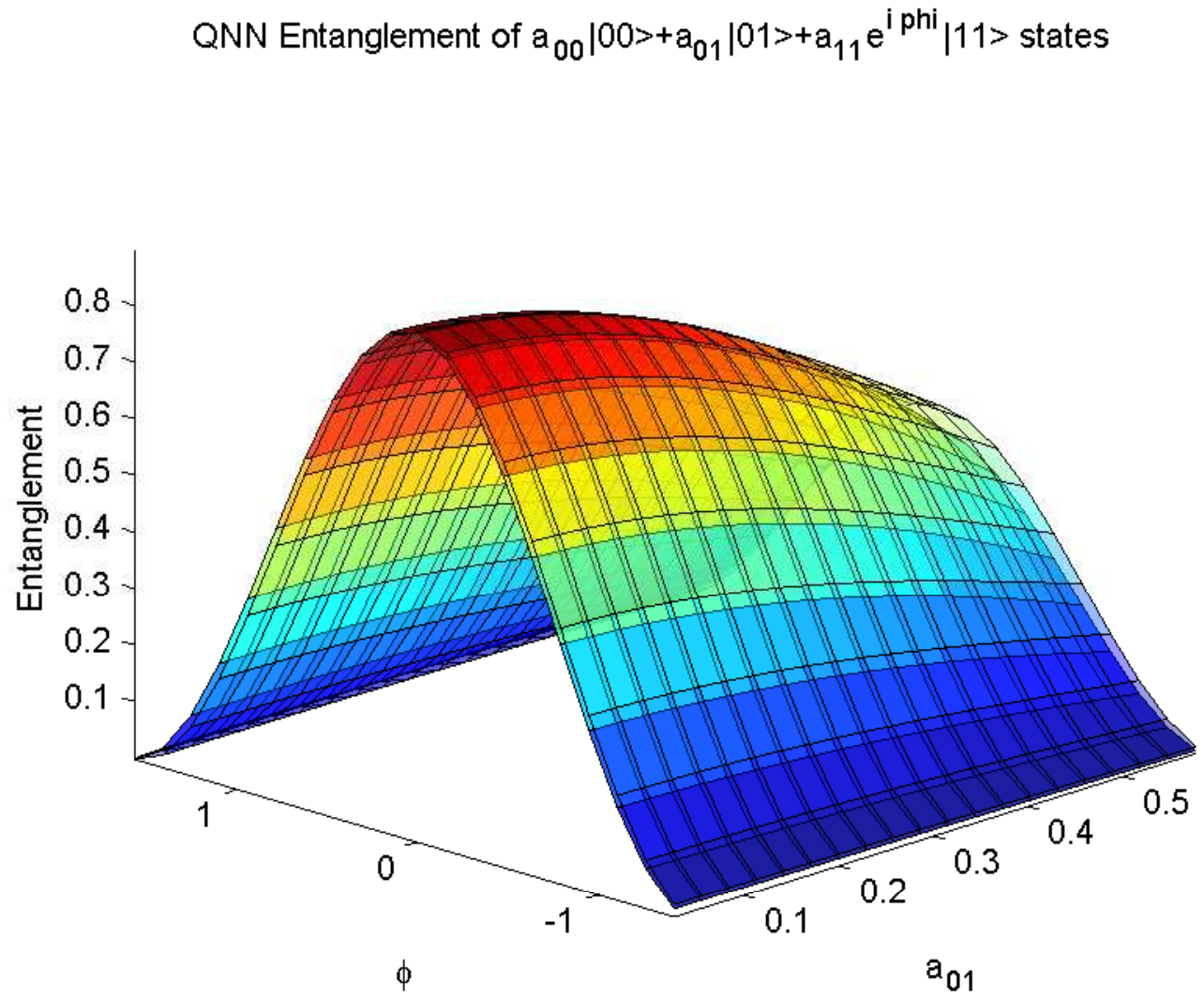}
\caption{The entanglement of $[a_{00}|00\rangle + a_{01}|01\rangle + e^{i \phi}a_{11}|11\rangle]$, as a function of $\phi$ and $a_{01}$, as calculated by the QNN,  and compared with a curvefit to $0.9\cos^{2}(1.3a_{01})\cos^{2}(\phi)$ (the partially transparent grid, superimposed.) \label{qnn_p_model_jan12fig}}
\end{figure}

The phase information seems inextricably linked with the entanglement information. For the simple case of a fully entangled state, either $\frac{1}{\sqrt{2}}[|00\rangle + e^{i \phi}|11\rangle]$ or $\frac{1}{\sqrt{2}}[|01\rangle + e^{i \theta}|10\rangle]$, our phase indicator returns the cosine squared (of $\phi$ or $\theta$, respectively) for the projections, so the procedure is straightforward: Using two copies of a given state, measure the phase offset on one copy, then apply a rotation operator to the other, and measure the entanglement.  Results are shown in Figure \ref{fixbellfig}.
The linkage does happen in both directions, though. If we examine the anomalous oscillation in, e.g., Figure \ref{QNNBellfig}, it is also remarkably close to a cosine squared. A good approximation is shown in Figure \ref{qnn_model_approximation_jan12fig}. Results are exactly similar for the $[a_{01}|01\rangle + e^{i \theta}a_{10}|10\rangle]$ states. For partially entangled states of the type $[a_{00}|00\rangle + a_{01}|01\rangle + e^{i \phi}a_{11}|11\rangle]$ (and symmetrical equivalents, such as $[a_{00}|00\rangle + a_{01}|10\rangle + e^{i \theta}a_{10}|10\rangle]$ ) we also get a function of the ``contaminant'' (for the above state, the magnitude $a_{01}$) times the cosine squared. This is shown in Figure \ref{qnn_p_model_jan12fig}.

\section{Discussion and further work}

We have shown that, for the simple case of a single phase offset of a pure state, we can train a quantum neural network to correct the anomalous oscillation in an entanglement witness. If we know that our experimental setup produces, say, pairs of states of the form $\frac{1}{\sqrt{a_{00}^{2} + a_{11}^{2} } } [a_{00}|00\rangle+e^{i \phi}a_{11}|11\rangle]$ with a high degree of probability, our method would allow for good reproducible determination of the entanglement. Of course this is not a complete solution even to the measurement of the entanglement of a two qubit system. That is, our results in the previous section do not transform Figure \ref{cxcoeffig} completely into Figure \ref{realfig}, as we have allowed only one of the coefficients at a time to be complex. 

We consider this work to be only proof-of-concept: Though our phase indicator does give us all relative phases, we need something more to solve this problem completely, for three reasons. First, because each measurement we make on a quantum system destroys all other information, so that we would need, minimum, four copies of our input state. With so many measurements necessary, it seems it might be easier simply to determine the state completely and then to calculate $E_{F}$ or whatever we wish to know, analytically. Second, inversion of the phase indicator target functions is increasingly problematic for the general input state. And third, it is not true that if we successfully unwind all the phases we will be able to recover the entanglement of the original state. Consider the entanglement of formation of the ``flat state'' (equal magnitudes of all of the charge basis states $\frac{1}{2}[|00\rangle+e^{i \xi}|01\rangle+e^{i \theta}|10\rangle+e^{i \phi}|11\rangle]$). The concurrence for this state is simply $C = |\sin[\frac{\phi-(\xi + \theta)}{2}]|$, and since $E_{F}$ is a monotonic bijective function of the concurrence, we can easily see that (since $C$ is maximum) $E_{F} = 1$ whenever the difference in phase between the ``Bell part'' and the ``EPR part'' is $\pi$. That is, if the ``Bell-type'' entanglement and the ``EPR-type'' entanglement are exactly in phase, the superposition is a product state, and, thus, has zero entanglement; if they are exactly out of phase, the density matrix cannot be factorized, and, thus, is entangled. 
This type of phase dependence actually causes entanglement, so, if we wish to determine the entanglement for a general input state, we will have to do more than simply determine all the phases and unwind them. It is not yet completely clear how much of this behavior the trained QNN net has already captured, since for \underline{all} real coefficients, including negative ones, we get good results (Figure \ref{realfig}).  Further work is needed, and is ongoing.


We have not considered here the multiple qubit systems that are of most interest to practical quantum computing; however, it is worth considering that a neural approach has natural advantages for scale-up. In fact, our entanglement indicator did easily generalize from two to three to four to five qubits \cite{nabic, eb12}.  We have also not considered mixed states, though our entanglement indicator \cite{eb08} did work also for these types of states, which gives us hope that this, larger, work may be easily extended to those also.
\ifCLASSOPTIONcaptionsoff
  \newpage
\fi



%

%

\begin{IEEEbiographynophoto}{E.C. Behrman}
earned an Sc.B from Brown University in mathematics, and an M.S. in chemistry and Ph.D. in physics from University of Illinois at Urbana-Champaign, where she performed research using Feynman path integrals to study the dynamics of tunneling systems in condensed phases at finite temperatures.  She did postdoctoral work at SUNY Stony Brook, and worked for four years at NYS College of Ceramics at Alfred University, doing research in calcium aluminosilicate glasses, chalcogenide glasses, viscosity of glass melts, and high temperature ceramic superconductors. She is currently professor of mathematics and of physics at Wichita State University where she has taught for twenty-two years.  She teaches undergraduate and graduate courses in physics at all levels, including quantum mechanics, statistical mechanics, and mathematical methods. Her work at Wichita State includes inorganic buckyballs and buckytubes, reaction intermediates, and defects in radiation-damaged materials as well as quantum neural computing.
\end{IEEEbiographynophoto}

\begin{IEEEbiographynophoto}{R.E.F. Bonde}
earned B.S. degrees in Aerospace Engineering, Physics (honors), and Mathematics (honors) from Wichita State University, where, in addition to quantum neural networks, he also did research with the Pierre Auger Array. He is currently a Ph.D. graduate student at the University of Texas at Arlington, working on the ATLAS experiment. 
\end{IEEEbiographynophoto}

\begin{IEEEbiographynophoto}{J.E. Steck}
has a B.S. and M.S. in Aerospace Engineering and a Ph.D. in Mechanical Engineering from the University of Missouri-Rolla where he performed research in the use of finite element methods in mechanics, fluids and aero-acoustics. He has done postdoctoral work in artificial neural networks as control systems for the Intelligent Systems Center at the University of Missouri-Rolla, and worked for two years at McDonnell Aircraft Company in the Aerodynamics department doing flight dynamics support for flight simulation, wind tunnel and flight testing of the AV-8B aircraft.  He is currently professor of Aerospace Engineering at Wichita State University where he has taught for seventeen years.  He teaches undergraduate and graduate courses in flight dynamics and controls, artificial neural networks, and computational methods.   His current work includes: intelligent adaptive control systems for general aviation aircraft, integrated aircraft structural health monitoring, quantum neural computing, optical aircraft ice detection, optical neural computing, and the use of artificial neural networks for system modeling and control.
\end{IEEEbiographynophoto}

\begin{IEEEbiographynophoto}{J.F. Behrman}
will earn a B.A. in physics (honors) from Harvard University in May 2013. In addition to work in quantum neural computing, she has also done research in the history of physics, under the direction of H. Georgi of Harvard; simulation and analysis of cosmic ray showers, under the direction of P. Horowitz of Harvard, and simulation and experimental work on disordered InO at the superconductor-insulator transition, under the direction of D. Shahar of the Weizmann Institute. 
\end{IEEEbiographynophoto}







\end{document}